\documentclass[a4paper,twocolumn,11pt,unpublished]{quantumarticle}
\pdfoutput=1
\usepackage[utf8]{inputenc}
\usepackage[english]{babel}
\usepackage[T1]{fontenc}
\usepackage{amsmath}
\usepackage{hyperref}

\usepackage{tikz}
\usepackage{lipsum}
\usepackage{mathrsfs}
\usepackage{amsfonts}
\usepackage{amssymb}
\usepackage{graphicx}
\usepackage{amsthm}
\usepackage{dsfont}
\PassOptionsToPackage{compress}{natbib}
\usepackage[numbers]{natbib}
\usepackage{subfigure}
\usepackage{textcomp}
\usetikzlibrary{shapes,arrows}
\usepackage{cancel}
\usepackage{makecell}
\usepackage{soul}

\newcommand{\figdir}{.}

\newcommand{\be}{\begin{equation}}
\newcommand{\ee}{\end{equation}}
\newcommand{\ket}[1]{\vert{ #1 }\rangle}

\newcommand{\ketbra}[2]{\vert #1 \rangle \langle #2 \vert}

\newcommand{\pool}{\mathcal{P}}

\begin{document}

\title{How Much Entanglement Do Quantum Optimization Algorithms Require?}

\author{Yanzhu Chen}
\affiliation{Center for Quantum Information Science \& Engineering, Virginia Tech, Blacksburg, VA 24061, U.S.A.}
\affiliation{Department of Physics, Virginia Tech, Blacksburg, VA 24061, U.S.A.}
\author{Linghua Zhu}
\affiliation{Department of Chemistry, University of Washington, Seattle, WA 98195, U.S.A.}
\author{Chenxu Liu}
\affiliation{Center for Quantum Information Science \& Engineering, Virginia Tech, Blacksburg, VA 24061, U.S.A.}
\affiliation{Department of Physics, Virginia Tech, Blacksburg, VA 24061, U.S.A.}
\author{Nicholas J. Mayhall}
\affiliation{Center for Quantum Information Science \& Engineering, Virginia Tech, Blacksburg, VA 24061, U.S.A.}
\affiliation{Department of Chemistry, Virginia Tech, Blacksburg, VA 24061, U.S.A.}
\author{Edwin Barnes}
\affiliation{Center for Quantum Information Science \& Engineering, Virginia Tech, Blacksburg, VA 24061, U.S.A.}
\affiliation{Department of Physics, Virginia Tech, Blacksburg, VA 24061, U.S.A.}
\author{Sophia E. Economou}
\affiliation{Center for Quantum Information Science \& Engineering, Virginia Tech, Blacksburg, VA 24061, U.S.A.}
\affiliation{Department of Physics, Virginia Tech, Blacksburg, VA 24061, U.S.A.}

\date{July 9, 2023}

\maketitle

\begin{abstract}
Many classical optimization problems can be mapped to finding the ground states of diagonal Ising Hamiltonians, for which variational quantum algorithms such as the Quantum Approximate Optimization Algorithm (QAOA) provide heuristic methods. 
Because the solutions of such classical optimization problems are necessarily product states, it is unclear how entanglement affects their performance. 
An Adaptive Derivative-Assembled Problem-Tailored (ADAPT) variation of QAOA improves the convergence rate by allowing entangling operations in the mixer layers whereas it requires fewer CNOT gates in the entire circuit. 
In this work, we study the entanglement generated during the execution of ADAPT-QAOA. Through simulations of the weighted Max-Cut problem, we show that ADAPT-QAOA exhibits substantial flexibility in entangling and disentangling qubits. 
By incrementally restricting this flexibility, we find that a larger amount of entanglement entropy at earlier stages coincides with faster convergence at later stages. 
In contrast, while the standard QAOA quickly generates entanglement within a few layers, it cannot remove excess entanglement efficiently. 
Our results demonstrate that the role of entanglement in quantum optimization is subtle and provide guidance for building favorable features into quantum optimization algorithms.
\end{abstract}

\maketitle

\section{Introduction}

In today's quantum devices, decoherence sets limits on the circuit depth. Quantum-classical hybrid variational algorithms, which bypass the need for deep quantum circuits, show promise for providing advantage over purely classical computing~\cite{Peruzzo2014, Farhi2014, Farhi2016, McClean2016, Cerezo2021_1}. In these algorithms, a parametrized ansatz state is realized through a parametrized quantum circuit, on which suitable measurements are carried out to provide information for classical optimization. 
A suitably chosen ansatz is the key to ensure the correct solution is reached or well approximated. 
Some approaches, such as the hardware-efficient ansatz, allow the ansatz the capacity of realizing any possible unitary operator in the Hilbert space~\cite{Kandala2017, Sim2019}. 
On the other hand, a highly expressive ansatz may lead to difficulty in optimization~\cite{McClean2018, Cerezo2021, Arrasmith2021, Holmes2022}. 
This has prompted a search for more problem-specific algorithms~\cite{Wecker2015, Grimsley2019, Hadfield2019, Choquette2021}. 
To this end, an Adaptive Derivative-Assembled Problem-Tailored (ADAPT) protocol has been developed~\cite{Grimsley2019, Tang2021, Shkolnikov2021}. First proposed to find the ground state energies of molecules, the ADAPT protocol constructs the ansatz iteratively, leading to a shallow quantum circuit and fast convergence. Its flexibility also allows one to incorporate some information about the hardware into the ansatz.

The application of hybrid variational algorithms extends beyond simulating physical and chemical systems. In general, optimization problems are of paramount importance to essentially every industry. While it is not yet clear whether quantum computers will offer significant speedups to optimization problems, the huge economic and technological repercussions of such an achievement have motivated the community to explore quantum advantage in the context of optimization. Entanglement has long been considered to play an essential role in quantum computing, although a precise statement in each specific context demands caution~\cite{Jozsa2003, Gross2009, Woitzik2020}. 
The role of entanglement is especially mysterious in quantum optimization algorithms, where both the initial state and the target state are unentangled. Nevertheless, it is often believed that creating some entanglement can provide an advantage to such algorithms. 
Many classical optimization problems, such as graph coloring and traveling time optimization, can be mapped to finding the ground state of an Ising Hamiltonian, where hybrid variational algorithms can be naturally applied~\cite{Lucas2014, Kochenberger2014, Hadfield2017, Oh2019, Shaydulin2019, Tabi2020, Bravyi2022, Weggemans2022}. 

A well-known heuristic algorithm for these problems is the Quantum Approximate Optimization Algorithm (QAOA)~\cite{Farhi2014}. It seeks the solution by optimizing the expectation value of the problem Hamiltonian with respect to an ansatz that alternates between a unitary evolution of the problem Hamiltonian and another unitary that amounts to a simultaneous $X$ rotation of all the qubits. The generator of the latter is called a mixer operator. 
Although the ansatz shares similarities with a trotterized version of an adiabatic evolution, its performance is limited by problem features, locality, and symmetry~\cite{Zhou2020, Akshay2020, Bravyi2020, Farhi2020, Farhi2020_1}. 
Moreover, analyses of computational resources and experimental requirements for QAOA point out difficulties in its near-term application~\cite{Guerreschi2019, Harrigan2021, Weidenfeller2022}.

To improve the performance of QAOA, one can modify it with the ADAPT approach~\cite{Zhu2020, Warren2022}. This algorithm, dubbed ADAPT-QAOA, builds up the ansatz layer by layer. While preserving the alternating structure of QAOA, it replaces the mixer at each step with an operator selected based on the energy gradient. 
It was observed that convergence was drastically improved in the Max-Cut problem when two-qubit operators were included in the mixer operator pool~\cite{Zhu2020}. Such two-qubit operators provide additional entangling operations in the ansatz. 
The possibility of having them in the mixer layers lends ADAPT-QAOA more freedom in entanglement production while the standard QAOA can only entangle the qubits through the unitary generated by the Hamiltonian.  
Despite the fact that ADAPT-QAOA is free to insert more entangling operations in the ansatz, somewhat counterintuitively the final ansatz it produces requires \textit{fewer} CNOT gates overall compared to the standard QAOA for a given convergence threshold~\cite{Zhu2020}. This indicates that the role of entanglement in quantum optimization algorithms is subtle and raises the question of how much entanglement is really needed to improve convergence.

Depending on the formalism chosen for the computation, entanglement can be seen as the key feature that sets quantum computing apart from classically simulable processes~\cite{Ekert1998, Jozsa2003}. 
On the other hand, excessive entanglement in certain settings can make classical optimization harder~\cite{Patti2021, Marrero2021}. 
Examining entanglement in the context of optimization problems can shed light on the efficiency and favorable features of hybrid variational algorithms. 
After simulating non-interacting fermions with the hardware efficient ansatz, Ref.~\cite{Woitzik2020} concluded that entanglement of an undesirable type can be a hindrance rather than a resource. In the Hamiltonian variational ansatz, entanglement entropy is observed to be between the level required in the ground state and the level in a random state~\cite{Wiersema2020}. In classical optimization problems, however, the target states do not intrinsically require a finite amount of entanglement. This makes such problems great testing grounds for studying whether existence of entanglement helps accelerate the optimization process. Ref.~\cite{Diez-Valle2021} suggests that for quadratic unconstrained binary optimization problems a product state ansatz may outperform entangled ones under certain conditions.

In this work, we investigate the role entanglement plays in the performance of ADAPT-QAOA for the weighted Max-Cut problem, which falls in the family of quadratic unconstrained binary optimization problems. 
The amount of entanglement entropy at each layer of the algorithm is calculated and compared to that produced by the standard QAOA. 
From this we find that a larger amount of entanglement entropy does not guarantee a higher convergence rate. 
Our simulations reveal that ADAPT-QAOA has substantial flexibility in entangling and disentangling qubits. 
By comparing its performance across different levels of flexibility, we find that a larger amount of entanglement entropy at earlier stages of ADAPT-QAOA can lead to faster convergence at later stages. 
We also find that ADAPT-QAOA favoring entangling mixer operators converges faster on average, although the first few optimized states doe not generate more entanglement in this case.  
The rest of the paper is organized as follows. In Sec.~\ref{sec:simulation} we describe the details of our simulations, with the results summarized in Sec.~\ref{sec:result}. We conclude and discuss the implications of our work in Sec.~\ref{sec:conclusion}.


\section{Numerical simulation}
\label{sec:simulation}

To study the entanglement generated in the process of variational algorithms, we consider the problem of finding the ground state of the Hamiltonian
\be
	H_{\rm MaxCut} = \frac{1}{2}\sum_{i<j} w_{ij} (Z_iZ_j - I),
\ee
where $\{w_{ij}\}$ are the edge weights of a graph, $Z_i$ is the Pauli $Z$ operator acting on the qubit labelled by $i$, and $i, j$ run from $0$ to $n-1$. This is equivalent to the weighted Max-Cut problem. 
Since the identity term in $H_{\rm MaxCut}$ only contributes a constant shift to the objective function and a global phase to the ansatz, we drop it and use the cost Hamiltonian 
\be
	H = \frac{1}{2}\sum_{i<j} w_{ij} Z_iZ_j
\ee
both as the objective function and in the ansatz. 
All the computational states are eigenstates of $H$. Since any two states related by the symmetry $F=\otimes_i X_i$ share the same energy, there are at least two degenerate ground states. In the standard QAOA, with the reference state, the Hamiltonian, and the mixer operator all respecting the symmetry $F$, the solution will always be a symmetric superposition of computational basis states. In ADAPT-QAOA, the selection criterion that determines the mixer operator at each layer ensures that only symmetric operators get selected~\cite{Zhu2020}. Consequently, the solution is also a symmetric superposition. 
In this work, we focus on the role of entanglement in the performance of the algorithm. To eliminate the effect caused by the entanglement required in the target state, 
we take either of the following two approaches. In the first we measure one of the qubits before calculating the entanglement entropy, which corresponds to projecting the state of the qubit to $0$ or $1$ in the simulation. We refer to this as the symmetry-preserving case. The second is directly breaking the symmetry of the Hamiltonian by adding to $H$ a small perturbation $fZ_0$, where $f$ is a small positive constant. The 2-fold ground space degeneracy is lifted, and the target ground state is a unique product state unless an accidental degeneracy is present. In this approach the Hamiltonian is
\be
	\tilde{H} = fZ_0 + \frac{1}{2}\sum_{i<j} w_{ij} Z_iZ_j,
\ee
which we refer to as the symmetry-breaking case.

We simulate the performance of both the standard QAOA and ADAPT-QAOA for 6-vertex degree-5 and 8-vertex degree-7 weighted regular graphs. In each case, 512 instances are generated in which the edge weights are randomly drawn from the discrete set $\{0.1, 0.2, \dots, 0.9\}$. We set $f=0.05$ so that the symmetry breaking term is always smaller than the energy gaps. To accommodate the symmetry-breaking setting, we choose a symmetry-breaking reference state in ADAPT-QAOA. Specifically, the state of qubit $0$ is initialized to $\vert1\rangle$ while the other qubits are initially in the state $\vert+\rangle$. 

The pool $\pool$ from which the mixer operators are selected for ADAPT-QAOA contains more options than the original mixer operator $M$ for the standard QAOA. For a $n$-qubit system, the pool is
\begin{align}
	\pool = & \{ M=\sum_{i=0}^{n-1} X_i, N = \sum_{i=0}^{n-1} Y_i\} \nonumber \\
	& \cup \{X_j, Y_j\}_{j =0, \dots, n-1} \nonumber \\
	& \cup \{X_jX_k, Y_jY_k, X_jY_k, X_jZ_k, Y_jZ_k \}_{j,k =0, \dots, n-1, j\neq k}. 
\end{align}
This is a complete pool, according to Ref.~\cite{Tang2021}, in that the ADAPT-QAOA ansatz built from this pool can approximate any unitary to arbitrary precision given sufficiently many layers. In Ref.~\cite{Zhu2020}, a significant improvement in performance is observed for ADAPT-QAOA with this pool compared to standard QAOA. Note that in the symmetry-breaking case we restore the operators that anticommute with the operator $F$. At the $l$-th layer, the operator $e^{-i\beta_lA_l} e^{-i\gamma_l\tilde{H}}$ is added to the optimal state at the $(l-1)$-th layer, where the mixer operator $A_l$ is chosen from $\pool$ after initializing $\gamma_l$ to some small non-zero value $\gamma_0$ and $\beta_l$ to $0$~\cite{Zhu2020}. In this work we set $\gamma_0=0.01$. The selection of the mixer operators takes a greedy approach and follows the ADAPT criterion: Among the operators in the pool, choose the one that maximizes the magnitude of the derivative of the energy (or more generally, the objective function) with respect to the new variational parameter $\beta_l$~\cite{Grimsley2019, Tang2021, Zhu2020}. With the $l$-th mixer fixed, \textit{all} the $2l$ variational parameters in the current ansatz are then (re-)optimized, where the previous $2(l-1)$ parameters are initialized to the optimal values obtained in the previous layer. We set the total number of layers to $p=15$ for the 6-vertex graphs and to $p=20$ for the 8-vertex graphs.

There have been various strategies for initializing the variational parameters in the standard QAOA~\cite{Shaydulin2019, Zhou2020, Sack2021, Li2021, Zhang2021}, and an initialization strategy based on optimal parameters for fewer layers has been shown to outperform random initialization~\cite{Zhang2021}. Here, we take an initialization strategy similar to that for ADAPT-QAOA: At the $l$-th layer, the $2(l-1)$ parameters that appear in the $(l-1)$-layer ansatz are initialized to the previously optimized values, while the two new parameters in front of $\tilde{H}$ and $M$ are initialized to $\gamma_0$ and $0$, respectively. 

In summary, the ansatze in ADAPT-QAOA, ADAPT-QAOA in the symmetry breaking setting, and the standard QAOA are, respectively,
\begin{align}
	&\vert\psi_{\rm AQ}\rangle = \left( \prod_{l=1}^p e^{-i\beta_lA_l} e^{-i\gamma_l\tilde{H}} \right) \otimes \overset{n}{\underset{i=1}{\bigotimes}} \ket{+}_i, \label{eq:A-Q} \\
	&\vert\psi_{\rm AQsb}\rangle = \left( \prod_{l=1}^p e^{-i\beta_lA_l} e^{-i\gamma_l\tilde{H}} \right) \ket{1}_0 \otimes \overset{n-1}{\underset{i=1}{\bigotimes}} \ket{+}_i, \label{eq:A-Q-sb} \\
	&\vert\psi_{\rm QAOA}\rangle = \left( \prod_{l=1}^p e^{-i\beta_lM} e^{-i\gamma_l\tilde{H}} \right) \overset{n-1}{\underset{i=0}{\bigotimes}} \ket{+}_i, \label{eq:QAOA}
\end{align} 
For the optimal state produced at each layer, we calculate the entanglement entropy for two types of bipartitions. One is the entanglement entropy across a middle cut separating half of the qubits from the rest. This measure overlooks the entanglement within either set of qubits, but any possible bias introduced by this particular bipartition will be mitigated after averaging over graphs with randomly generated edge weights. 
The other is the entanglement entropy between each qubit and the remaining qubits, averaged over all qubits. This measure is $0$ if and only if the system is in a product state. By taking the entanglement entropy as the measure, we do not distinguish different entanglement classes~\cite{Coffman2000, Wei2003, Virmani2001, Plenio2007, Horodecki2009}.

To see if entanglement is favorable when the ansatz is adaptive, we examine the behavior of ADAPT-QAOA in different settings. Since ADAPT-QAOA has better performance in the symmetry-preserving case than in the symmetry-breaking case, we take the former for this simulation. First we vary the level of flexibility for ADAPT-QAOA. If a quantum device has limited connectivity, it is costly to realize an entangling gate between the qubits that are not directly connected. This is common in quantum processors such as superconducting devices, where one can adjust the mixer operator pool accordingly. We compare the full pool with two restricted pools containing two-qubit operators for the nearest neighbors on a ladder configuration (a $2\times3$ grid for the $6$-qubit case and $2\times4$ grid for the $8$-qubit case) and on a linear configuration, respectively.

Then we rescale the magnitude of the energy gradient in each ADAPT step to increase (decrease) the chance that ADAPT-QAOA adds an entangling mixer operator to the ansatz. Specifically, for the two-qubit operators in the pool, we multiply the energy gradient by a factor of $1-\delta$ where $0\leq\delta<1$. Compared with the previous case where extra entangling operations are introduced as a byproduct of the gradient-based adaptive procedure, this rescaling directly influences the ansatz construction based on the entangling feature. We choose four values $\delta\in\{\pm0.1, \pm0.5\}$ in our simulation.

In addition to entanglement entropy, we also calculate the entanglement spectrum, which reveals more information about a state~\cite{Li2008}. 
We know that $\pool$ is a complete pool for ADAPT-QAOA, and the controllability of the standard QAOA can be quantified by the dimension of the dynamical Lie algebra generated by $\tilde{H}$ and $M$~\cite{Larocca2021, Zhang2021}. 
With a number of layers that does not grow with the system size, neither ansatz has the capacity to reach all possible states in the Hilbert space. This is a desired feature, since one would face the challenge of barren plateaus in optimizing a sufficiently generic ansatz~\cite{McClean2018, Cerezo2021, Larocca2021}.  
To numerically test how close the distribution of ansatz circuits is to that of random unitaries, we turn to the entanglement spectrum, which has been employed to study the properties of the Hamiltonian variational ansatz~\cite{Wiersema2020}. For a state $\ket{\psi}$ describing a system partitioned into subsystems A and B, it is defined as the spectrum of the operator $-\log(\rho_{\rm A})$, where 
\be
	\rho_{\rm A}={\rm Tr}_{\rm B}(\ketbra{\psi}{\psi}).
\ee 
At each layer, it is calculated for the ansatz state with the variational parameters taking random values. 

Throughout this work, the minimization is implemented with the Nelder-Mead method in the SciPy library~\cite{scipy}. The codes are derived from those in Refs.~\cite{Grimsley2019, Zhu2020} and are available \href{https://github.com/yzchen-phy/adapt-vqe}{here}.


\section{Results}
\label{sec:result}

\subsection{Comparing ADAPT-QAOA with the standard QAOA}

\begin{figure*}[!ht]
	\includegraphics[width=1.0\linewidth]{\figdir/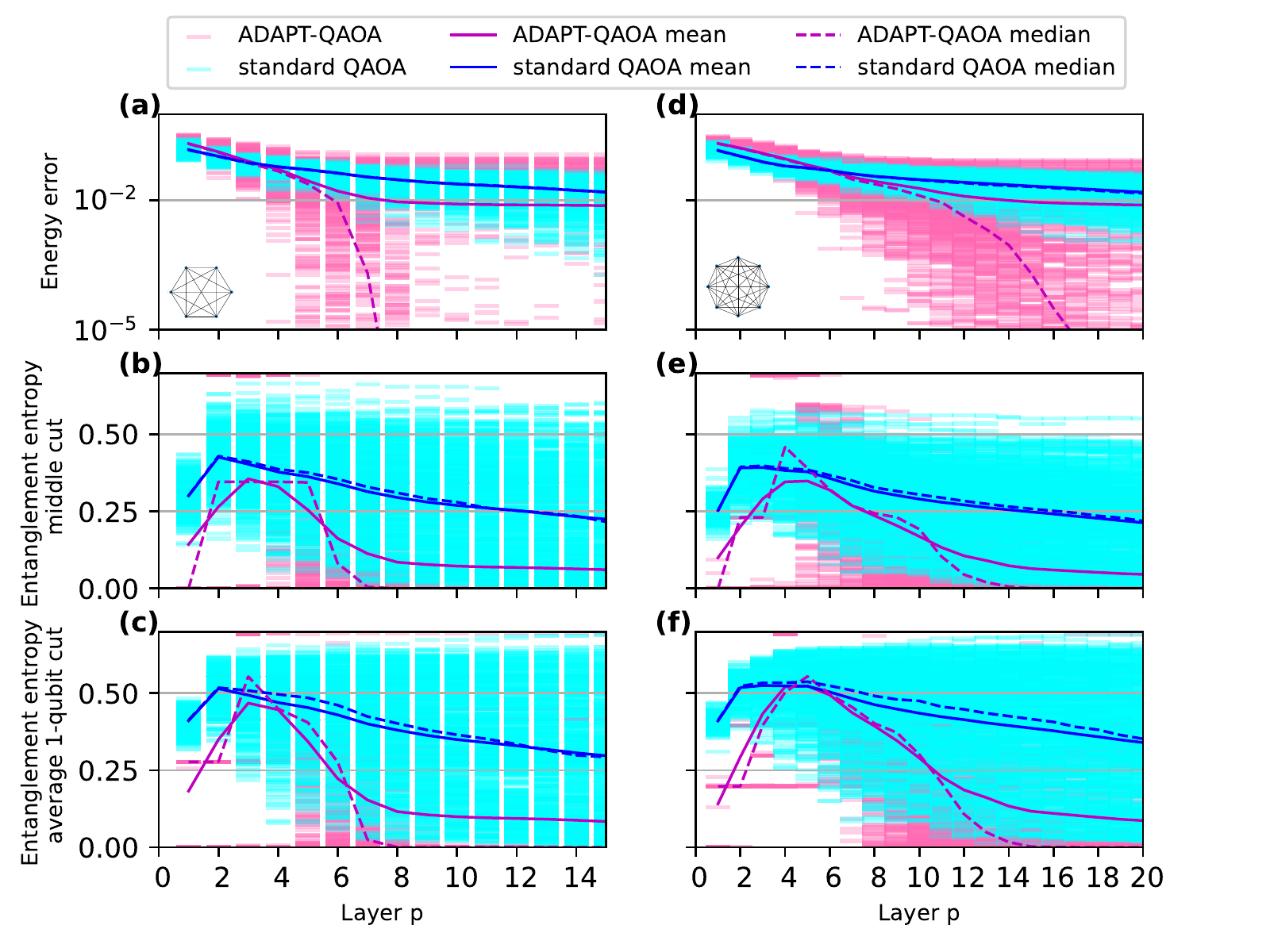}
	\caption{Energy error normalized to the MaxCut value (a, d), entanglement entropy across a middle cut (b, e), and the average entanglement entropy across 1-qubit cuts (c, f) for the optimized state with each number of QAOA layers, in the symmetry-preserving case. The panels on the left (a, b, c) and right (e, f, g) are for the 6-qubit and 8-qubit graphs, respectively. The performance of ADAPT-QAOA (magenta) is compared to that generated by the standard QAOA (blue). The scattered short lines are from different graph instances while the solid and dashed lines are the mean and the median of them, respectively.}
	\label{fig:reg_sym}
\end{figure*}

\begin{figure*}[!ht]
	\includegraphics[width=1.0\linewidth]{\figdir/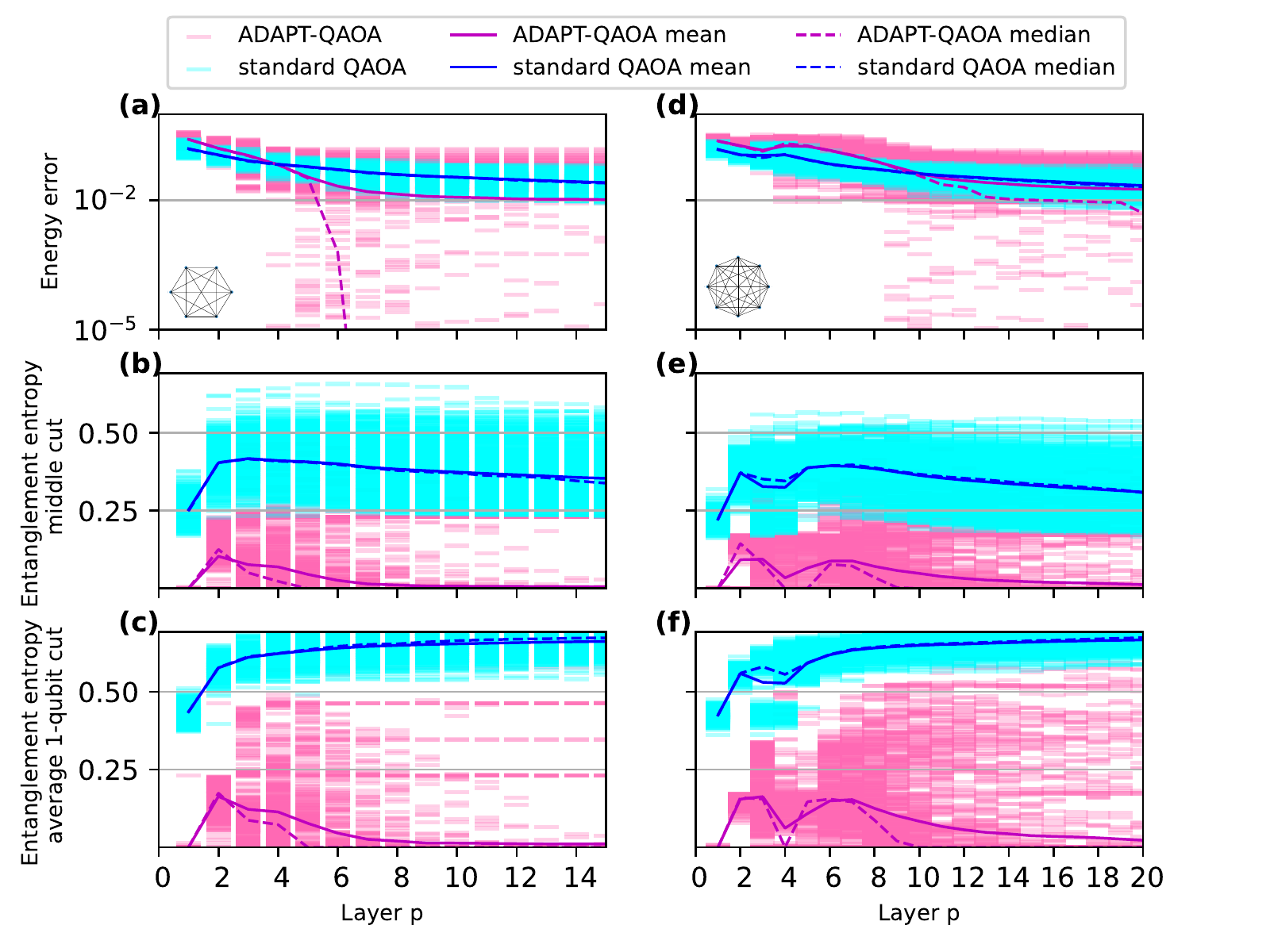}
	\caption{Energy error normalized to the MaxCut value (a, d), entanglement entropy across a middle cut (b, e), and the average entanglement entropy across 1-qubit cuts (c, f) for the optimized state with each number of QAOA layers, in the symmetry-breaking case. The panels on the left (a, b, c) and right (d, e, f) are for the 6-qubit and 8-qubit graphs, respectively. The performance of ADAPT-QAOA (magenta) is compared to that generated by the standard QAOA (blue). The scattered short lines are from different graph instances while the solid and dashed lines are the mean and the median of them, respectively.}
	\label{fig:reg_sb}
\end{figure*}

\begin{figure*}[!ht]
	\includegraphics[width=1.0\linewidth]{\figdir/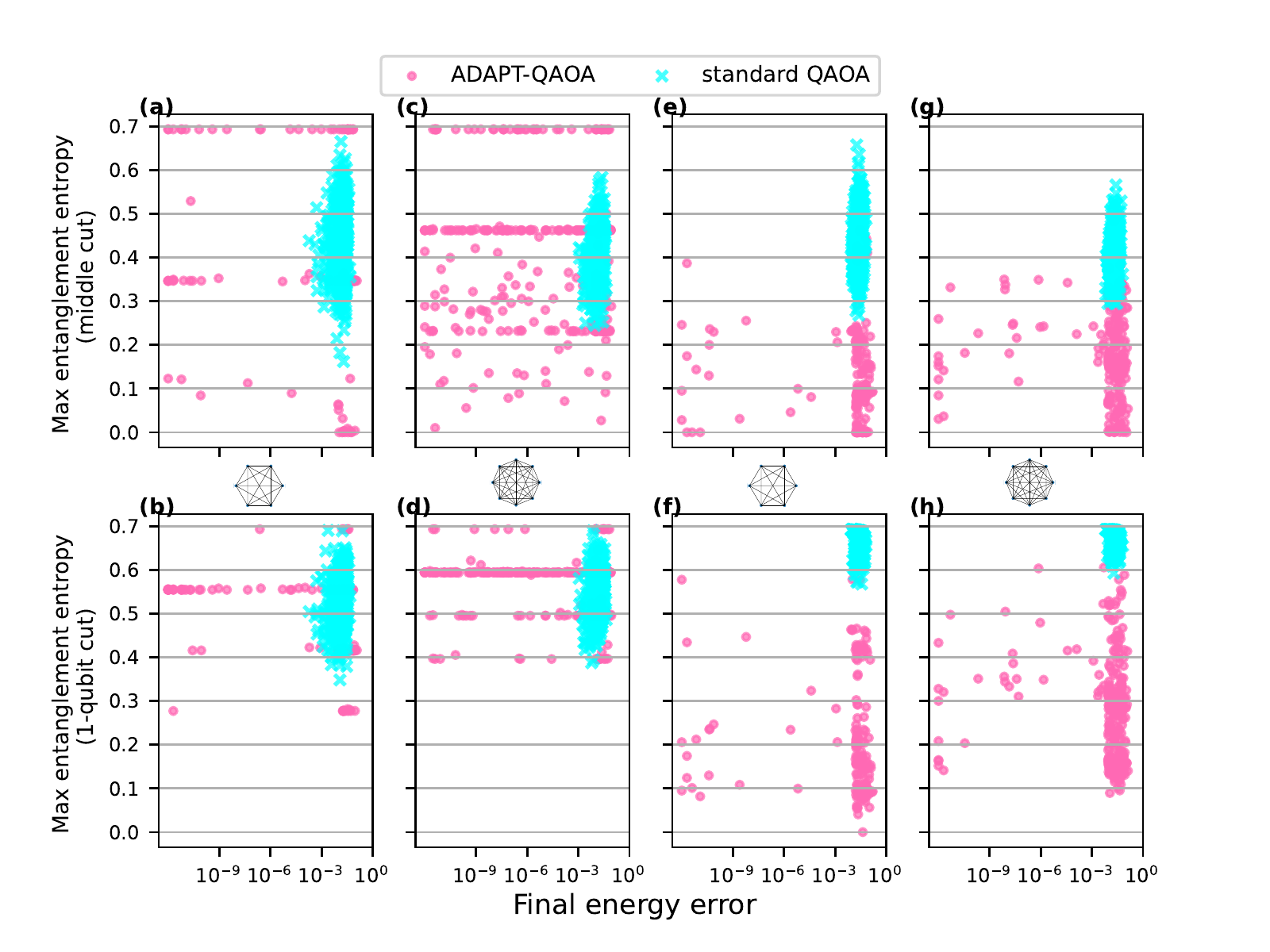}
	\caption{Maximum entanglement entropy in the intermediate optimized states as a function of the final energy error in the symmetry-preserving case (a, b, c, d) and the symmetry-breaking case (e, f, g, h). The performance of ADAPT-QAOA (magenta dot) is compared to that generated by the standard QAOA (blue cross). Each point is from one graph instance.} 
	\label{fig:reg-err-ent}
\end{figure*}

\begin{figure}[!ht]
	\includegraphics[width=1.0\linewidth]{\figdir/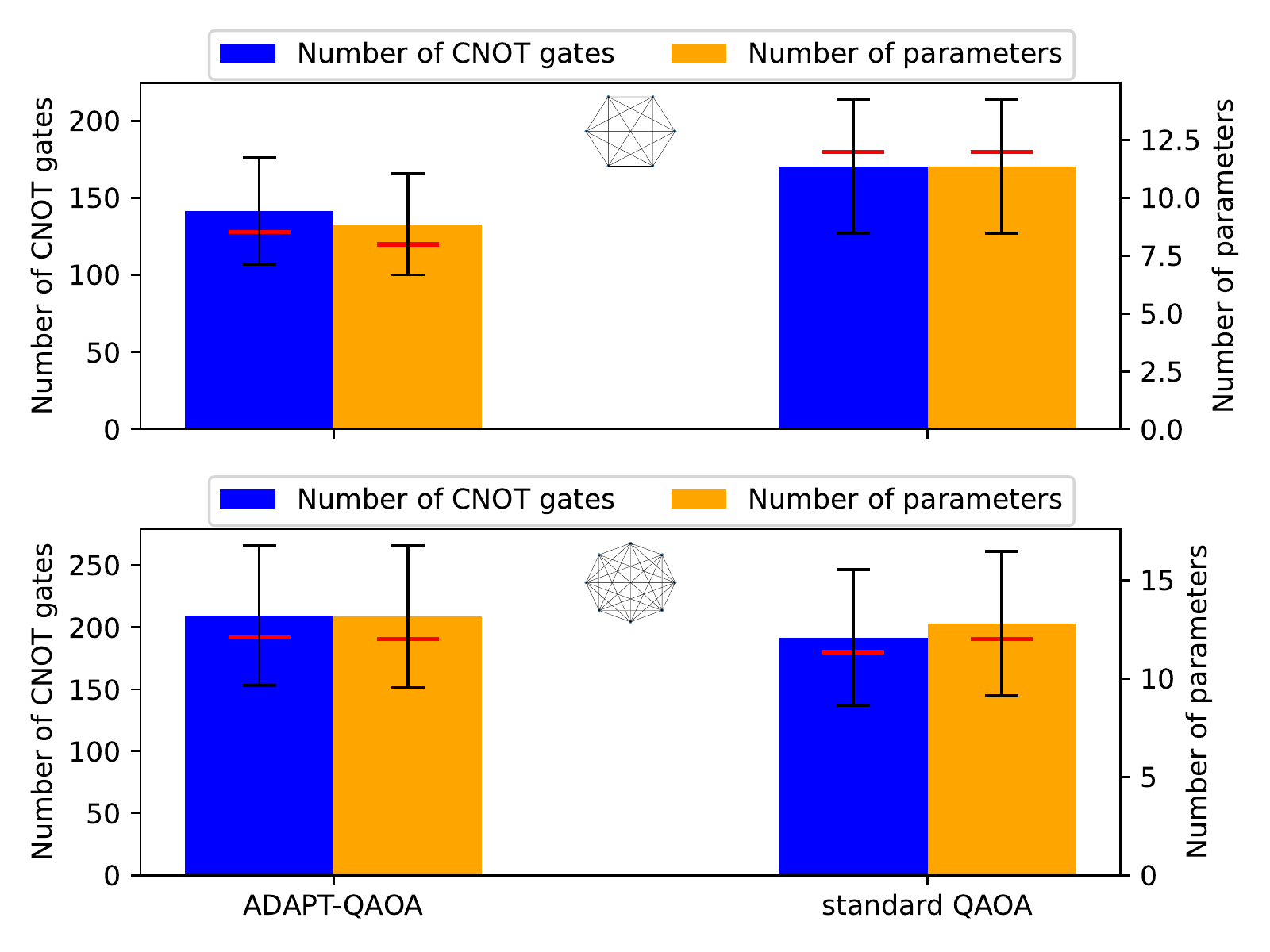}
	\caption{Number of CNOT gates (left, blue) and number of variational parameters (right, orange) required by ADAPT-QAOA and the standard QAOA with a symmetric reference state, when the energy error normalized to the MaxCut value reaches the threshold $0.05$ in the symmetry-preserving case. The error bars show the standard deviation. Panels (a) and (b) are for the 6-qubit and 8-qubit graphs, respectively. The numbers are undercounted due to the cutoff at 15 (20) layers for the 6-qubit (8-qubit) graphs.}
	\label{fig:resource}
\end{figure}

As shown in Figures~\ref{fig:reg_sym} and \ref{fig:reg_sb}, with very few layers ADAPT-QAOA finds an entangled optimized state. As the number of layers increases, the entanglement entropy in the solution drops down close to zero as required for the ground state. 
In contrast, the entanglement entropy of the optimal state found by the standard QAOA increases fast in the first few layers but decreases more slowly in the later layers. 
As shown in Fig.~\ref{fig:reg_sym}, the ADAPT-QAOA ansatz can generate more entanglement at early stages than the standard QAOA ansatz for some graph instances, with the effect of symmetry removed. 
As shown in Fig.~\ref{fig:reg_sb}, the entanglement in the standard QAOA ansatz remains high throughout the algorithm, largely due to the symmetry. In this case, the symmetry-breaking ADAPT-QAOA generates less entanglement while still outperforming it as the number of layers increases.  

Fig.~\ref{fig:reg-err-ent} shows the final energy error as a function of the maximum entanglement entropy generated by the optimal state at any layer. We observe that there is no definite relation between the achieved accuracy and the maximum entanglement. In other words, creating more entanglement does not necessarily accelerate the optimization process in generic quantum optimization algorithms.

Both cases demonstrate that the additional flexibility in ADAPT-QAOA allows it to control the rise and fall of entanglement to its advantage. With few layers, both algorithms are capable of finding solutions with more entanglement than required. With more layers, however, ADAPT-QAOA can remove the excess entanglement far more efficiently than the standard QAOA. We attribute this to the two-qubit mixer operators in the ADAPT-QAOA ansatz. Their presence helps with both the production and the removal of entanglement in the state whereas the Hamiltonian is the only operation in the standard QAOA ansatz that can do so. 

Fig.~\ref{fig:resource} shows the number of CNOT gates and the number of variational parameters required in the ansatz circuit when the energy error gets below some threshold. It has been shown that for classical approximate algorithms it hard to get an approximation ratio better than $16/17\approx94.1\%$~\cite{Hastad2001}. We set the threshold to be $0.05$, below this value. 
Among the 512 graph instances, the energy error has not reached the threshold at the last layer of ADAPT-QAOA for 29 (16) instances of the $6$-qubit ($8$-qubit) graphs. These corresponds to the cases where ADAPT-QAOA converges to an excited state and increasing the number of layers does not lower the energy further. We exclude these cases when calculating the average resources.


\subsection{Comparing different levels of flexibility in ADAPT-QAOA}

\begin{figure*}[!ht]
	\includegraphics[width=1.0\linewidth]{\figdir/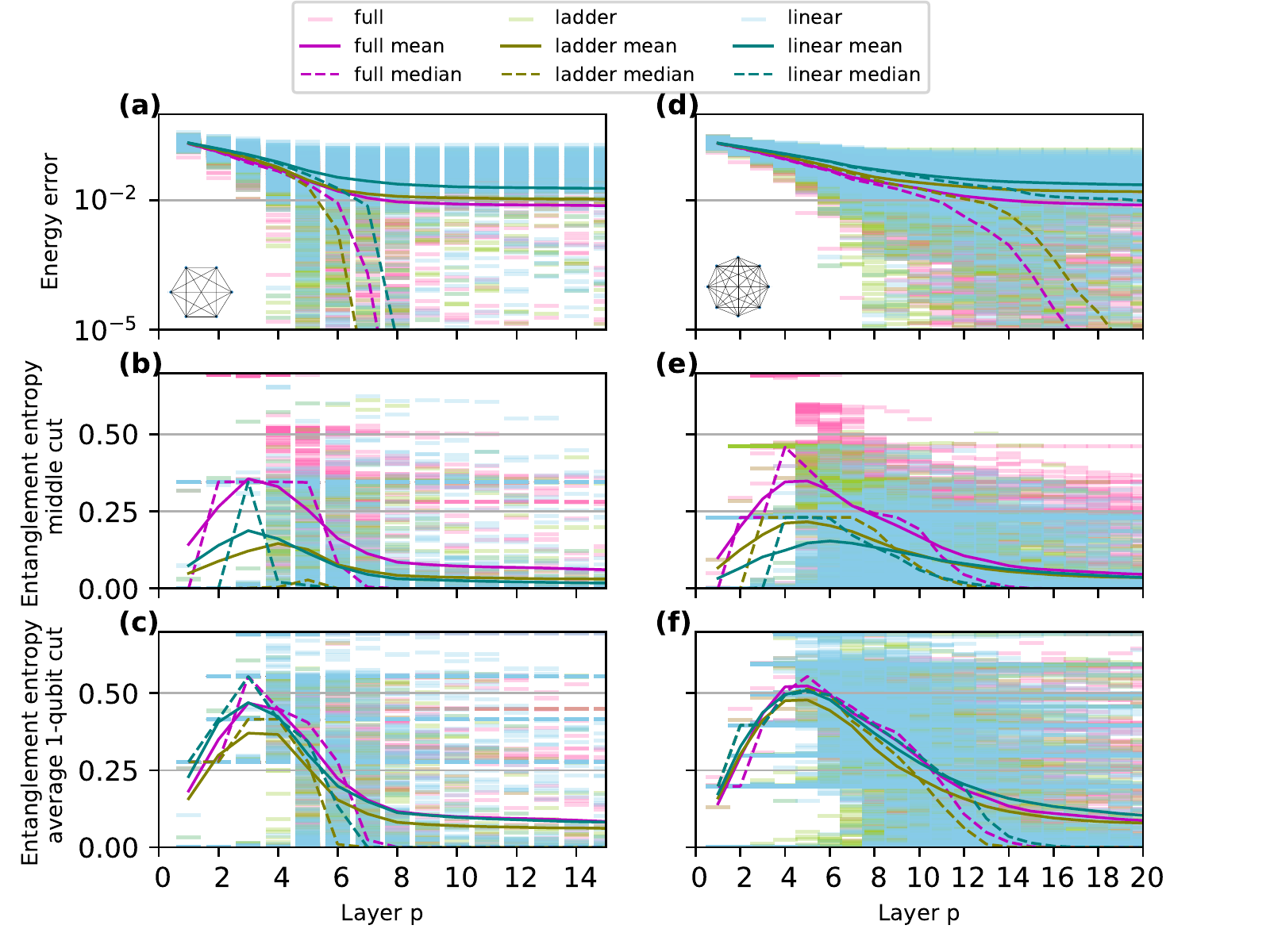}
	\caption{Energy error normalized to the MaxCut value (a, d), entanglement entropy across a middle cut (b, e), and the average entanglement entropy across 1-qubit cuts (c, f) for the optimized state with each number of QAOA layers, in the symmetry-preserving case. The panels on the left (a, b, c) and right (d, e, f) are for the 6-qubit and 8-qubit graphs, respectively. The performance of ADAPT-QAOA with a full operator pool (magenta) is compared to that with potential mixer operators are restricted to the nearest neighbors on a ladder configuration (olive) or a linear configuration (teal). The scattered short lines are from different graph instances while the solid and dashed lines are the mean and the median of them, respectively.} \label{fig:con}
\end{figure*}

The observation that ADAPT-QAOA is able to generate an appropriate amount of entanglement can be reinforced by examining its performance as we constrain its flexibility to varying degrees. With fewer options for entangling operations, ADAPT-QAOA exhibits slower convergence when averaging over graphs with different edge weights.  
One feature accompanying slower convergence is less efficient entanglement generation, as shown in Fig.~\ref{fig:con}. In the first few layers, the energy expectation values of the optimal states are close to each other regardless of the operator pool. Despite similar energies, ADAPT-QAOA with full connectivity produces more entanglement entropy than with reduced connectivity. With more layers, it is able to disentangle more quickly and get close to the target state whereas reduced connectivity leads to a solution with more entanglement entropy than needed. In this case, better overall performance coincides with more entanglement entropy at earlier stages of the algorithm, an indication that more entanglement than required in the target state can be helpful. 


\subsection{Comparing ansatze favoring or penalizing entanglement}

\begin{figure*}[!ht]
	\includegraphics[width=1.0\linewidth]{\figdir/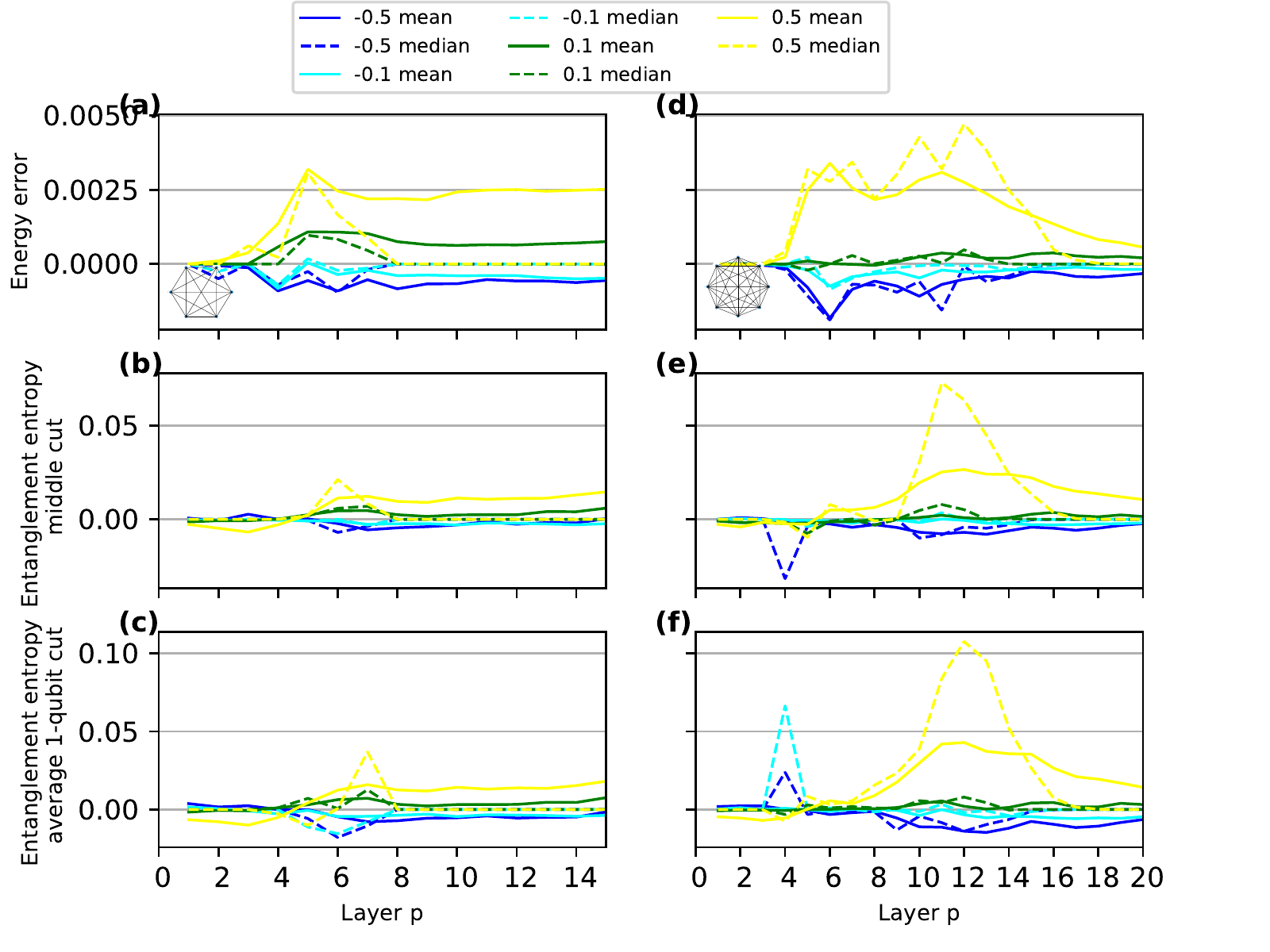}
	\caption{Difference between the ADAPT-QAOA optimized state with a penalty $\delta$ and that without the penalty, in energy error (a, d), entanglement entropy across a middle cut (b, e), and the average entanglement entropy across 1-qubit cuts (c, f) at each number of QAOA layers in the symmetry-preserving case. The panels on the left (a, b, c) and right (d, e, f) are for the 6-qubit and 8-qubit graphs, respectively. We omit the individual graph instances but retain the mean (solid line) and the median (dashed line) only.} \label{fig:pen}
\end{figure*}

Having seen that ADAPT-QAOA, given a higher level of flexibility, tends to generate more entanglement at early stages, we now explore whether manually boosting entanglement is beneficial. We focus on the difference between the values obtained by ADAPT-QAOA with a penalty term $\delta$ and those obtained without the penalty, averaged over different graph instances, as shown in Fig.~\ref{fig:pen}. In the first 3-4 layers ADAPT-QAOA reaches similar energy errors across different $\delta$'s. As the number of layers increases further, a negative $\delta$ favoring entangling operators outperforms a positive $\delta$ penalizing entangling operators on average. At these layers, the optimized states associated with $\delta<0$ has less entanglement entropy than those associated with $\delta>0$. This again underlines the importance of the ability to disentangle the state. We remark that the reward or penalty introduced by rescaling the energy gradient is different from adding fixed entangling operations to the ansatz. It still compares different entangling operations based on the energy gradient criterion while it is unclear which entangling operations would be useful in the latter case.


\subsection{Entanglement spectrum}

\begin{figure*}[!ht]
	\includegraphics[width=1.0\linewidth]{\figdir/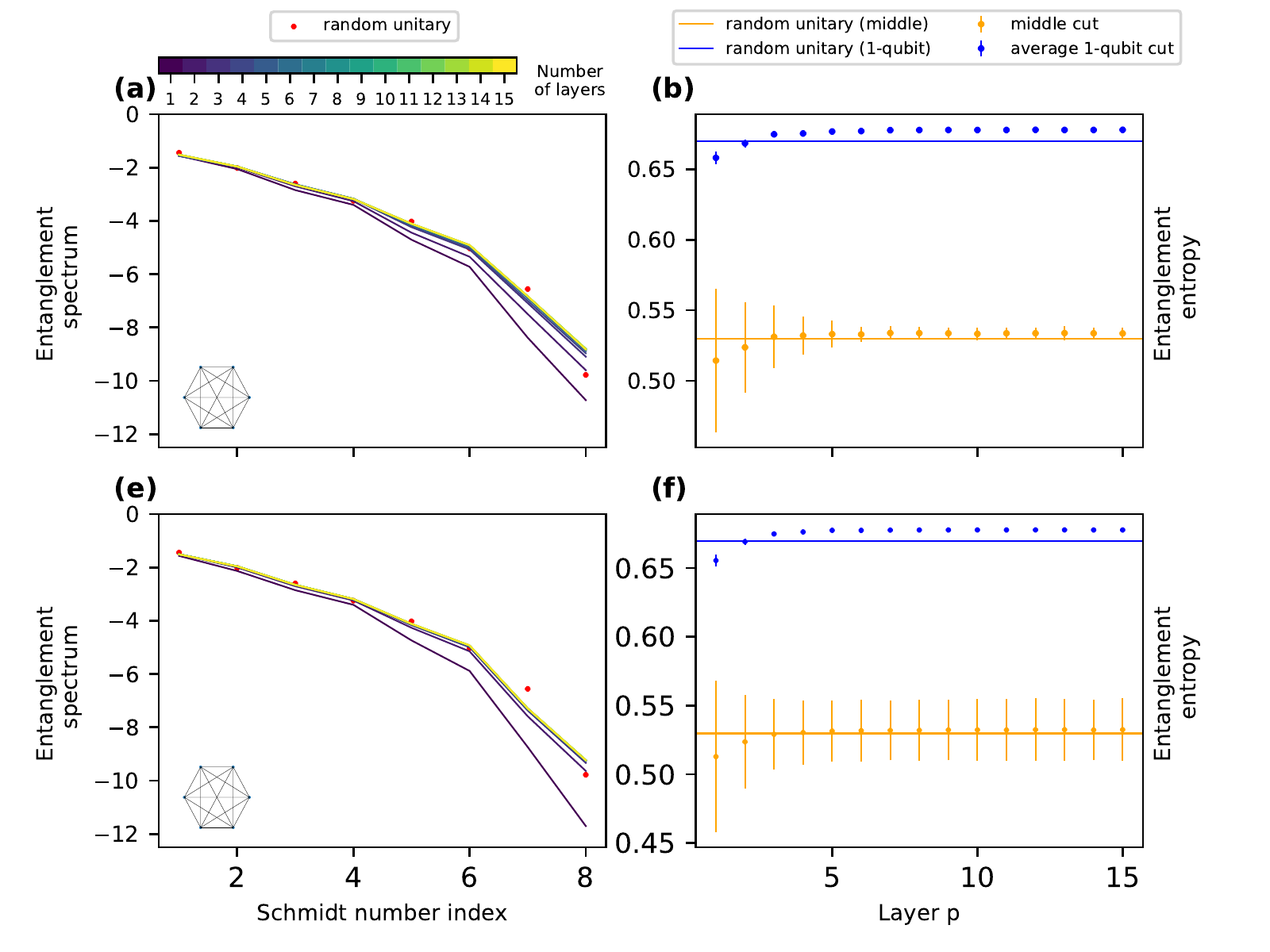}
	\caption{Entanglement spectrum, entanglement entropy across a middle cut, and the average entanglement entropy across 1-qubit cuts for the ADAPT-QAOA ansatz (a, b) and the standard QAOA ansatz (c, d), respectively, with the variational parameters randomly sampled between $0$ and $20\pi$ in the symmetry-preserving case. We average over 50 $6$-qubit graph instances and the error bars on the entanglement entropy represent the standard deviation among the graph instances.} 
	\label{fig:RSspec}
\end{figure*}

With the ability to take any operators from the pool $\pool$, ADAPT-QAOA can in principle generate all possible unitaries if there are no restrictions on the number of layers. For each problem, the operator selection process ensures that the ansatz is specific. 
As the number of layers increases, the ansatz becomes more expressive. 
We compare the ansatze produced at each layer by ADAPT-QAOA and the standard QAOA. 
In both cases, we replace the optimized variational parameters in the ansatz with values randomly drawn from $[0, 20\pi)$, and average over $1000$ such states for the ansatz associated with one graph instance. This offers an indirect probe into how general the possible states for a chosen ansatz can be. 

In Fig.~\ref{fig:RSspec}, we present the entanglement spectrum across a middle cut given each number of layers, averaged over 50 $6$-qubit graph instances since the simulation is costly. 
Qualitatively, the entanglement spectrum becomes more similar to that produced by random unitaries as the number of layers increases for both ADAPT-QAOA and the standard QAOA. 
Interestingly, in terms of the average entanglement spectrum produced, the standard QAOA ansatz approaches the random unitary more quickly. 
This suggests that the ADAPT-QAOA ansatz is less generic, leading to easier optimization of the variational parameters.


\section{Conclusion and discussion}
\label{sec:conclusion}

In this work we investigate the role of entanglement in solving MaxCut problems with quantum optimization algorithms, where a finite amount of entanglement is not required in the target state. In the context of QAOA and its adaptive variation ADAPT-QAOA, we find that the algorithm performance can benefit from entanglement generation in the process. On the other hand, the relation between the convergence speed and the amount of entanglement in the intermediate state is not a simple positive correlation and it can be counterproductive to create too much entanglement.

In particular, by comparing the entanglement generated by ADAPT-QAOA at various stages of the algorithm to that generated by the standard QAOA, we observe that although the standard QAOA can quickly generate more entanglement entropy with only a few layers in the ansatz, it is inefficient in removing excess entanglement as the number of layers increases. Consequently, at later stages it does not perform as well as ADAPT-QAOA, in which both the Hamiltonian and the selected two-qubit mixer operators contribute to entanglement generation and removal. 
By examining ADAPT-QAOA with different mixer operator pools, we find that given the largest degree of flexibility, it produces more entanglement entropy at earlier stages and converges faster at later stages. This hints at the promise of entangled ansatze in quantum optimization algorithms. 
We stress that more entanglement entropy by itself does not imply faster convergence in the entire process. Rather, it is important to have suitable entangling operations. 
We also find that in ADAPT-QAOA, manually favoring entangling mixer operators in the ansatz improves the convergence on average.  
Finally, the average entanglement spectrum of the possible states for each ansatz suggests that with a given number of layers, ADAPT-QAOA is less expressive than the standard QAOA, which can reduce the difficulty in the classical optimization component of the variational algorithm.

These observations provide qualitative guidance in designing quantum optimization algorithms. 
Specifically, variational algorithms equipped with flexibility in entangling operations can outperform the algorithms with limited entangling operations.  
From this perspective, ADAPT-QAOA and the ADAPT approach in general demonstrate great potential in generating a suitable amount of entanglement for the optimization task. 
To offer better flexibility, the mixer operator pool should include multiqubit operators beyond those supported on nearest neighbors. 
It will thus be easier to demonstrate ADAPT-QAOA on platforms with less limitation on entangling operations between arbitrary pairs of qubits, such as trapped ion systems~\cite{Landsman2019, Zhang2020, Blumel2021} and Rydberg atoms~\cite{Levine2019, Graham2019, Graham2022}.

In our work, entanglement is quantified by the entanglement entropy associated with a bipartition, which does not reveal the many-body structure of the entanglement. 
Multipartite entanglement, where the number of parties grows with the system size, is necessary if the quantum algorithm is exponentially hard to simulate classically~\cite{Jozsa2003}. It thus remains to be investigated whether the heuristic quantum optimization algorithms studied here offer any advantage over classical algorithms. 

\begin{acknowledgements}
S. E. E. acknowledges support from the US Department of Energy (Award No. DE-SC0019318). E.B. and N.J.M. acknowledge support from the US Department of Energy (Award No. DE-SC0019199).
\end{acknowledgements}

\bibliographystyle{unsrtnat}
\bibliography{adapt_qaoa_ent}

\end{document}